\documentclass[prb,twocolumn,superscriptaddress,showpacs]{revtex4}

\usepackage{graphicx}

\newcommand{\etal}{{\em et al}.}
\newcommand{\EF}{$E_{\rm F}$}
\newcommand{\rt}{$2p\rightarrow3d$}
\newcommand{\vo}{V$_2$O$_3$}
\newcommand{\vcro}{(V$_{1-x}$Cr$_x$)$_2$O$_3$}
\newcommand{\vcrotwo}{(V$_{0.972}$Cr$_{0.028}$)$_2$O$_3$}
\newcommand{\vcroone}{(V$_{0.988}$Cr$_{0.012}$)$_2$O$_3$}
\newcommand{\vtio}{(V$_{1-x}$Ti$_x$)$_2$O$_3$}
\newcommand{\vtioone}{(V$_{0.99}$Ti$_{0.01}$)$_2$O$_3$}
\newcommand{\vmo}{(V$_{1-x}$M$_x$)$_2$O$_3$}
\newcommand{\stob}{$I_{s}/I_{b}$}

\begin{document}

\title{Photoemission study of (V$_{1-x}$M$_x$)$_2$O$_3$ (M=Cr, Ti)}

\author{S.-K. Mo}
\affiliation{Randall Laboratory of Physics, University of
Michigan, Ann Arbor, MI 48109}
\author{H.-D. Kim}
\affiliation{Pohang Accelerator Laboratory, Pohang 790-784, Korea}
\author{J. D. Denlinger}
\affiliation{Advanced Light Source, Lawrence Berkeley National
Laboratory, Berkeley, CA 94720}
\author{J. W. Allen}
\affiliation{Randall Laboratory of Physics, University of
Michigan, Ann Arbor, MI 48109}
\author{J.-H. Park}
\affiliation{Department of Physics, Pohang University of Science
and Technology, Pohang 790-784, Korea}
\author{A. Sekiyama}
\author{A. Yamasaki}
\author{S. Suga}
\affiliation{Department of Material Physics, Graduate School of
Engineering Science, Osaka University, 1-3 Machikaneyama,
Toyonaka, Osaka 560-8531, Japan}
\author{Y. Saitoh}
\affiliation{Department of Synchrotron Radiation Research, Japan
Atomic Energy Research Institute, SPring-8, Sayo, Hyogo 679-5143,
Japan}
\author{T. Muro}
\affiliation{Japan Synchrotron Radiation Research Institute,
SPring-8, Sayo, Hyogo 679-5143, Japan}
\author{P. Metcalf}
\affiliation{Department of Physics, Purdue University, West
Lafayette, IN 47907}

\date{Received \hspace*{30mm}}

\begin{abstract}
We present high-resolution bulk-sensitive photoemission spectra of \vmo\ (M=Cr, Ti). The measurements were made for the paramagnetic metal (PM), paramagnetic insulator (PI), and antiferromagnetic insulator (AFI) phases of \vmo\ with the samples of $x$ = 0, 0.012, and 0.028 for Cr-doping and $x$ = 0.01 for Ti-doping. In the PM phase, we observe a prominent quasiparticle peak in general agreement with theory, which combines dynamical mean-field theory with the local density approximation (LDA+DMFT). The quasiparticle peak shows a significantly larger peak width and weight than in the theory. For both the PI and AFI phases, the vanadium $3d$ parts of the valence spectra are not simple one peak structures. For the PI phase, there is not yet a good theoretical understanding of these structures. The size of the electron removal gap increases, and spectral weight accumulates in the energy range closer to the chemical potential, when the PI to AFI transition occurs. Spectra taken in the same phases with different compositions show interesting monotonic changes as the dopant concentration increases, regardless of the dopant species. With increased Cr-doping, the AFI phase gap decreases and the PI phase gap increases.
\end{abstract}


\maketitle

\section{\label{intro}Introduction}
\begin{figure}[!tb]
\includegraphics[width=3.3 in]{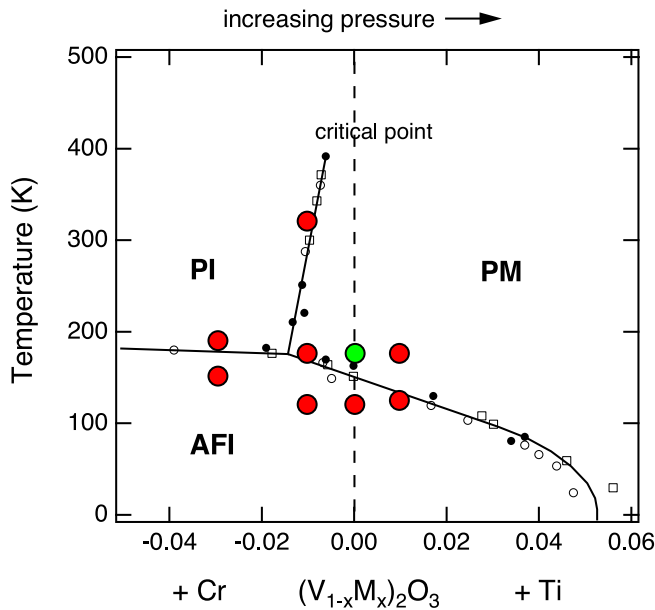}
\caption{\label{PD}(color online) Phase diagram of \vmo\ in $T$-$P$ or $T$-doping space. From Ref.\ [\onlinecite{McWhan}]. The green (light gray) and red (dark gray) dots indicate, respectively, the temperatures and doping levels of the samples for the PES data presented in our previous letter (Ref.\ [\onlinecite{Mo03}]) and those for the new data reported in this paper. }
\end{figure}

\vmo\ (M=Cr, Ti) has long been considered as a textbook example of the metal-insulator transition (MIT). It has a complex phase diagram with paramagnetic metal (PM), paramagnetic insulator (PI), and antiferromagnetic insulator (AFI) phases\cite{McWhan, Kuwamoto80} as presented in Fig.~\ref{PD}. The phase transitions are induced by changing temperature ($T$) and pressure, and by doping Cr or Ti in place of V. Pure \vo\ is in the PM phase for temperatures above $T_{N}\approx\ $155 K, below which it is in the AFI phase\cite{Moon70}. For Cr concentration $0.005 \lesssim x \lesssim 0.017$, the material undergoes a second transition, from PM to PI, with increasing $T$. The PM phase is completely suppressed for $x \gtrsim 0.017$, and then only a PI-AFI transition takes place. Ti-doping suppresses the AFI phase and makes the system PM at all $T$ for $x \gtrsim 0.05$. The PM and PI phases have a rhombohedral (corundum) structure, which becomes monoclinic in the low-$T$ AFI phase. The existence of a structural transition at the PM-AFI boundary suggests the possible importance of a changing lattice symmetry for the transition. In contrast, the PM to PI transition is isostructural, there being only a change of the lattice parameter $c/a$ ratio, which may imply that the transition is primarily of electronic origin. The PM to PI transition is generally considered to be a realization of the Mott-Hubbard (MH) MIT\cite{Mott, Gebhard}.

The MH scenario for the MIT in \vmo\ was originally put forth in the framework of the half-filled one-band Hubbard model, in which the on-site Coulomb repulsion ``$U$" competes with the site-to-site hopping ``$t$" that tends to make a broad band of bandwidth ``$B$". Even though the one-band Hubbard model may catch the essence of the MIT, it may oversimplify the actual electronic structure to such an extent that essential physics might well be lost. At issue has been the fact that the two $3d$ electrons of the V$^{3+}$ ion must be distributed to singly degenerate $a_{1g}$ and doubly degenerate $e^{\pi}_{g}$ orbitals derived from a trigonal distortion of the $t_{2g}$ manifold of the V $3d$ states and the reduction to the half-filled one-band Hubbard model consequently requires various assumptions\cite{Castellani}.

Recent experimental and theoretical development indeed show that multi-band realism should be considered to understand the electronic properties of \vmo. On the experimental side, analysis of polarized X-ray absorption spectroscopy\cite{Park00} has shown that the V$^{3+}$ spin is $S=1$, with large $e^{\pi}_{g}$ but also significant $a_{1g}$ occupations, invalidating the one-band model. The orbital occupation also shows abrupt changes at the phase boundaries of \vmo, suggesting that the orbital degrees of freedom play an important role in the MIT. On the theory side, models leading to $S=1$ have been developed\cite{Ezhov99, Mila00, Matteo02}, and a realistic description of the electronic structure including many-body dynamics was given by calculations that combine density functional theory in the local density approximation (LDA) with the dynamic mean field theory (DMFT) [LDA+DMFT].\cite{Held01, HeldLong, Keller04, Laad03} The LDA+DMFT calculations are consistent with $S=1$ and successfully describe the MIT betweem the PM and PI phases with realistic values of $U$.

LDA+DMFT calculations also provide detailed predictions for the V $3d$ single particle spectral functions, which can be measured by photoemission spectroscopy (PES). The most notable feature of the PES spectrum calculated by LDA+DMFT is a prominent quasiparticle (QP) peak at the Fermi energy (\EF) in the PM phase, which was not observed in previous PES measurements.\cite{Sawatzky79, Smith88, Smith94, Kim98, Schramme} We have reported in our previous letter\cite{Mo03} the observation of such a peak in the PM phase PES spectrum of \vo, showing general agreement as to the energies of spectral features, but significant differences in the peak width and weight, relative to the spectrum predicted for $T$ = 300~K in an LDA+DMFT calculation implemented using the quantum Monte Carlo (QMC) method. The experimental finding was enabled by using a small photon spot and high photon energy to enhance the bulk sensitivity of the spectrum.\cite{Maiti98, Sekiyama00, Sekiyama04}

In this paper we present new PES spectra in the PM, PI, and AFI phases of \vo, \vcro\ ($x$ = 0.012, 0.028), and \vtio\ ($x$ = 0.01) measured at different temperatures, indicated by the red (dark gray) dots in Fig.~\ref{PD}.  The data in Ref.\ [\onlinecite{Mo03}] was only for the green (light gray) dot in the figure.  Thus we now explore the changes in the spectra across the phase boundaries, the variation of the spectra with doping within a phase, the comparison of the spectra with each other and with recent theoretical findings, and the values and variations of the insulating phase gaps. We also show the origin of the need for a small photon spot, and we compare the results of extracting bulk spectra from takeoff angle variation and photon energy variation.  These topics are generally important for PES spectroscopy of correlated electron materials and our discussions and comparisons for these cleaved single crystal surfaces go beyond past treatments of this topic in the literature.

The paper is organized as follows: In Sec.~\ref{exp}, the experimental details of the PES measurements are described. Sec.~\ref{PMphase} presents our findings for the PM phase and gives our new results concerning surface sensitivity issues. PES spectra across the PM to PI, PM to AFI, and PI to AFI phase boundaries are presented and compared in Sec.~\ref{MIT} and discussed in terms of various theoretical calculations including DMFT. Sec.~\ref{Doping} compares the lineshapes of the PES spectra of the PM and AFI phases induced by varying dopant concentration. A summary is given in Sec.~\ref{Summary}, including a perspective on the insulator gap values.   Distortion of the V $3d$ lineshapes due to Auger emission in resonant PES (RESPES) is reported in the Appendix.

\section{\label{exp}Experimental}
Well-oriented single crystalline samples of \vo, \vcro\ ($x$ = 0.012, 0.028), and \vtio\ ($x$ = 0.01) were cleaved to expose a hexagonal (10$\bar{1}$2) plane. The cleaved samples were kept in a vacuum better than $2 \times 10^{-10}$~Torr. PES measurements at SPring-8 were made using circularly polarized photons with energy between $h\nu$ = $310$~eV and $700$~eV at the twin-helical undulator beam line BL25SU \cite{Saitoh00}. The Fermi level and overall energy resolution ($\approx$ 90~meV to 170~meV over the $h\nu$ range) were determined from the Fermi-edge spectrum of a Pd metal reference. PES measurements at the Advanced Light Source (ALS) were made at beamline 7.0, using photon energy between 200~eV and 520~eV. The overall resolution at the ALS was $\approx 350$~meV. The beam spot diameter was $\approx$ 170 $\mu$m at SPring-8\cite{footnote_beamspot} and $\approx$ 100 $\mu$m at the ALS.

The data were taken using the transmission mode of the SCIENTA SES200 (SPring-8) and SES100 (ALS) electron analyzers, for which the acceptance angle is about $\pm6$ degrees, which corresponds to $\pm1.2$ \AA$^{-1}$ in \textbf{k}-space at the photon energy $\approx$ 500~eV. Since the size of the first Brillouin zone of \vmo\ is less than 1.0~\AA$^{-1}$, our spectra are essentially \textbf{k}-integrated transverse to the analyzer axis. The temperature was controlled by an embedded resistive heater and a closed-cycle He cryostat. Surface integrity was well maintained under the vacuum and photon exposure, and continually monitored by checking the repeatability of the reference spectra. The phase transition temperatures are invariant to cycling through a transition and the spectra in each phase are repeatable within small variations in the relative intensities of the valence band peaks. These variations are consistent with the sample surface inhomogeneity described in Section~\ref{SS_SmallSpot} and small changes that occur in the measurement position during cycling.

The photon energy range used for the measurements includes the V \rt\ absorption edge ($h\nu \approx$ 517~eV) where a large resonance enhancement of the V $3d$ emission occurs. Therefore, one is tempted to use RESPES to obtain data with higher statistics in much less measurement time. Unfortunately, by comparing the spectra on and off resonance, we have found\cite{Mo04b} that Auger emission accompanies the RESPES spectra of all the phases of \vmo, making a small but significant distortion of the V $3d$ lineshapes. As a result, we have deliberately avoided the use of the photon energy around the resonance region. The details of the RESPES spectra near the V \rt~absorption edge and the Auger emission are presented in the Appendix.


\begin{figure}[!bt]
\includegraphics[width=3.3 in]{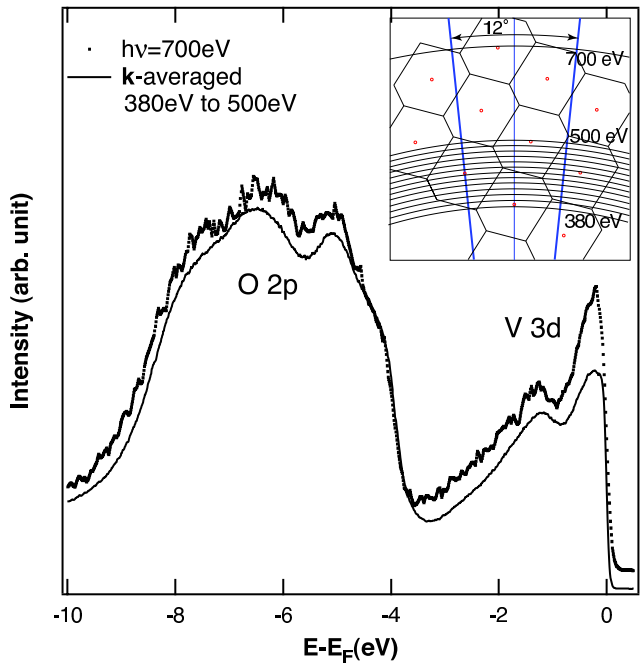}
\caption{\label{Overview}(color online) PES spectra of \vo\ taken with $h\nu$ = 700~eV and {\bf k}-averaged, from Ref.\ [\onlinecite{Mo03}]. The two curves are offset slightly for clarity. The inset shows the {\bf k}-space covered by the detector angular acceptance angle for various $h\nu$ on top of the Brillouin zone diagram of \vo\ in the PM phase.}
\end{figure}

\section{\label{PMphase}Bulk Sensitive spectrum \newline of the PM phase \vo~}

\subsection{Overview spectra}
The inset of Fig.~\ref{Overview} shows a cross section of the Brillouin zone in the high temperature rhombohedral phase of \vmo, for stacking normal to the cleavage plane, with the arcs in {\bf k}-space that are covered by the detector at various photon energies from 380~eV to 700~eV. Radial lines show the {\bf k} range corresponding to the analyzer acceptance angle of about $\pm 6^{\circ}$, which, as explained in Sec.~\ref{exp}, covers more than one Brillouin zone for any photon energy used in this study. We note that the loss of photoelectron momentum due to the incident photon momentum was not properly included in a similar figure that we presented previously\cite{Mo03}. The arcs in the inset now correctly reflect the effect of the photon momentum. The $h\nu$ = 700~eV spectrum of Fig.~\ref{Overview} shows the general character of the PM phase PES data of \vmo. These data are taken in the PM phase of \vo\ at $T$ = 175~K. The V $3d$ emission is well-separated from the O $2p$ emission, and the V $3d$ part of the spectrum shows a two-peak structure with a prominent \EF\ peak. We also show a {\bf k}-averaged spectrum obtained by summing spectra taken in 10~eV steps from 380~eV to 500~eV, thereby covering the whole Brillouin zone including the direction perpendicular to the sample normal. The \EF\ peak in the {\bf k}-averaged spectrum is smaller than that of the 700~eV spectrum, which is attributed to the reduced bulk sensitivity, rather than the variation caused by probing different {\bf k}-space arcs, as discussed in the following two subsections. 

\begin{figure}[!bt]
\includegraphics[width=3.3 in]{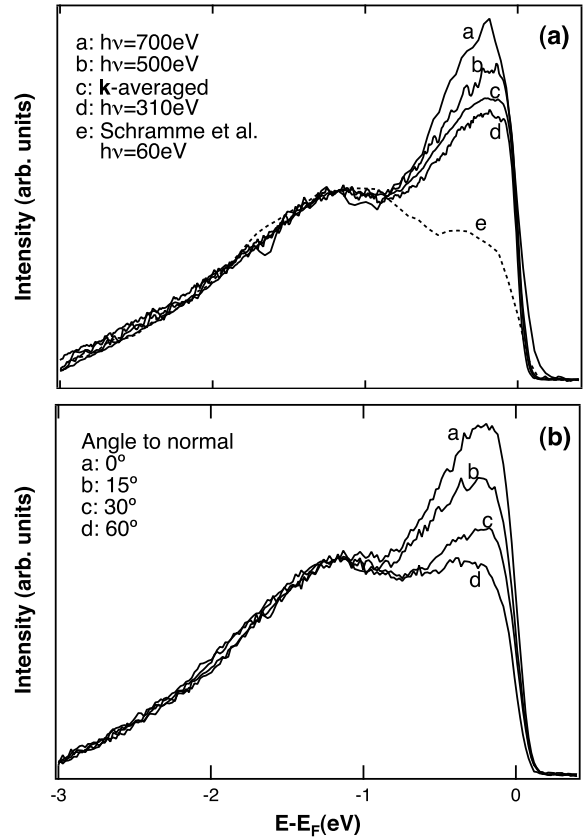}
\caption{\label{PhotonAngle} (a) PES spectra taken with varying $h\nu$, from Ref.\ [\onlinecite{Mo03}]. (b) PES spectra taken with varying emission angle from the surface normal ($h\nu$ = 500 eV). }
\end{figure}

\subsection{Bulk sensitivity}
With the assumption of an ideally smooth surface, the surface to bulk emission ratio \stob\ is given by\cite{Citrin83}, 
\begin{equation}
I_{s}/I_{b} = \exp(d/\lambda \cos\theta) - 1 \label{SvsB_eq}
\end{equation}
where $d$ is the thickness of the surface layer, $\lambda$ is the electron escape depth, and $\theta$ is the electron emission angle relative to the sample surface normal. To extract surface and bulk components of the PES spectrum, we have measured spectra with different \stob\ by changing the photon energy and the emission angle. The former relies on the fact that $\lambda$ varies strongly with the electron kinetic energy, and thus with the photon energy, and for the latter one utilizes the fact that the effective escape depth varies as the cosine of the angle from the normal in Eq.~(\ref{SvsB_eq}).

Fig.~\ref{PhotonAngle} shows V $3d$ spectra for varying (a) $h\nu$ and (b) the emission angle from the sample surface normal, respectively. A Shirley-type inelastic background\cite{Shirley72} has been removed in an identical way for each spectrum, such that the minimum between the V $3d$ and O $2p$ region becomes zero. In so doing we implicitly make a separation between V $3d$ and O $2p$ emission, which facilitates the comparison with the LDA+DMFT calculation that was made only for the V $3d$ electrons. The spectra are normalized over the range below -1~eV, rather than the area below the spectra, for ease of comparing the relative intensities of the \EF\ peaks. 

\begin{figure}[!tb]
\includegraphics[width=3.3 in]{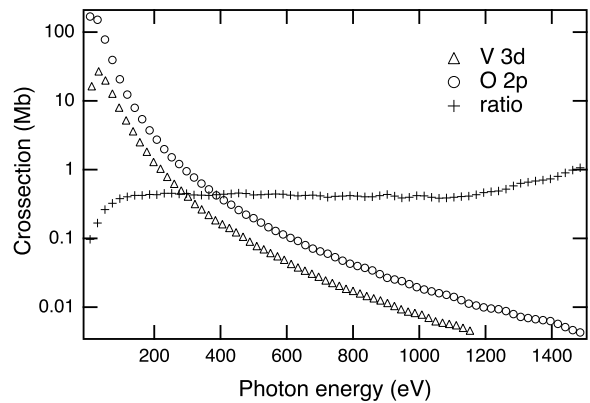}
\caption{\label{Crossection}Photo-ionization crossections\cite{YehLindau} for V $3d$ and O $2p$ in \vo, and their ratio. The ratio is nearly a constant for the photon energies used in this study.}
\end{figure}

Panel (a) of Fig.~\ref{PhotonAngle} shows $h\nu$ = 310~eV, 500~eV, and 700~eV spectra along with the {\bf k}-averaged spectrum taken from the previous subsection. Over the photon energy range used for the {\bf k}-averaging, 380 - 500 eV, the \EF\ intensity is never larger than that of the 500~eV spectrum. We also include 60~eV data taken from Ref.\ [\onlinecite{Schramme}], a high quality spectrum typical of low photon energy PES. The relative intensities of the \EF\ peaks in the spectra increase monotonically with increasing $h\nu$. This change in intensity cannot simply be attributed to the change in the photo-ionization cross-section as $h\nu$ varies. Fig.~\ref{Crossection} shows the change of the photo-ionization cross-section\cite{YehLindau} for V $3d$ and O $2p$ as the photon energy varies from 100~eV to 1500~eV, along with cross-section ratio. Proper multiplication factors are applied for V $3d$ and O $2p$ according to the number of atoms in a unit cell. The figure clearly shows that the ratio of the V $3d$ and O $2p$ cross-sections in \vo\ is nearly constant over the photon energy range used in this study, 300~eV to 700~eV, which shows that the change in the \EF\ intensity is not the result of a variation in the relative cross-sections of two peaks. Rather the change should be attributed to enhanced PES probe depth\cite{Gerken85} with increasing photon energy.

The importance of the probe depth is also shown in panel (b) by the monotonic decrease of the intensities of the \EF\ peaks as we change the emission angle away from the sample surface normal to 60$^{\circ}$, which effectively decreases the probe depth. The change with the angle variation of the relative heights of the \EF\ and -1.2~eV peaks bears a close similarity to the variation due to the photon energy change, to the degree that the 60$^{\circ}$ spectrum becomes very similar to the 60~eV spectrum. The variation of the relative intensity of the \EF\ peaks does not follow exactly the cosine dependence which one would expect from Eq.~(\ref{SvsB_eq}). This could be related to an inhomogeneous distribution of steps and roughness on the surface, which goes beyond the assumption of a smooth surface made in obtaining Eq.~(\ref{SvsB_eq}). The importance of such surface steps and roughness is explained in detail in the next subsection. Another possible source of the deviation from the simple cosine dependence is the elastic photoelectron scattering effect\cite{Jablonski99}. However, this effect is rather small for the small angles used in this study. Thus both the photon energy and angle dependences shown in Fig.~\ref{PhotonAngle} indicate increased bulk-sensitivity for the $h\nu$ = 700~eV spectrum. This conclusion is further supported by the extraction of the bulk spectrum, as explained below.

Using the formula for the electron escape depth from Ref.~[\onlinecite{Tanuma87}], we have obtained the values of $\lambda$ for the photon energies at which the data were taken. The value of $d$, 2.44\AA, is calculated from the crystal structure. $I_{b}$ and $I_{s}$ are uniquely determined for a given set of $\lambda$ and $\theta$, from Eq.~(\ref{SvsB_eq}) with the condition $I_{b} + I_{s} = 1$ required to preserve the total spectral weight. Then a PES spectrum can be written as a linear combination of bulk and surface spectra:
\begin{equation}
\label{SvsB_eq2}
\begin{array}{ccc}
    A_{1}(\omega) = I_{b1}B(\omega) + I_{s1}S(\omega)  &    \\
    A_{2}(\omega) = I_{b2}B(\omega) + I_{s2}S(\omega)  &   
\end{array}
\end{equation}
where $A_{1}(\omega)$ and $A_{2}(\omega)$ are two different PES spectra measured with different bulk and surface coeffcients $I_{b1}$, $I_{s1}$ and $I_{b2}$, $I_{s2}$, respectively. $B(\omega)$ and $S(\omega)$ are the bulk and surface spectra, respectively. Using Eq.~(\ref{SvsB_eq2}), surface and bulk spectra can be extracted from a pair of PES spectra measured with either different photon energies or different emission angles.

\begin{figure}[!tb]
\includegraphics[width=3.3 in]{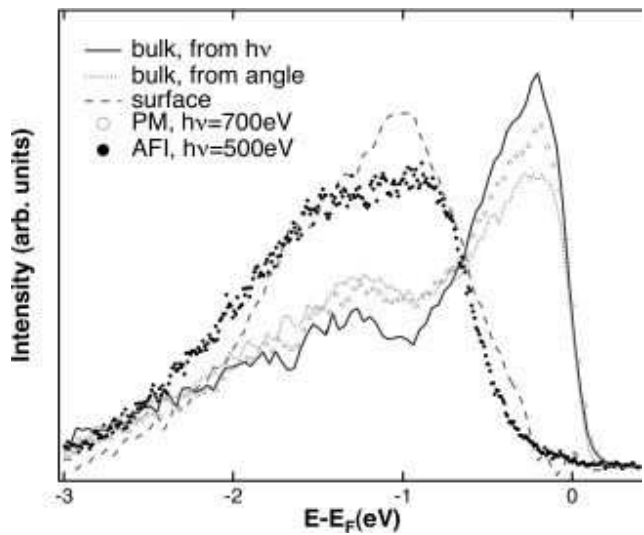}
\caption{\label{Bulk_Surface}(color online) Bulk and surface spectra of \vo\ extracted from various pairs spectra with differing photon energies and differing emission angles, thus having differing relative intensity of surface to bulk components. }
\end{figure}

Fig.~\ref{Bulk_Surface} shows the extracted bulk spectrum from a pair of spectra at photon energies 60~eV and 700~eV, and a pair of angle-dependent spectra for 15$^{\circ}$ and 60$^{\circ}$ at $h\nu$ = 500~eV. These results are not identical, but are similar to each other, and also to the 700~eV spectrum. Further, the bulk spectra extracted from other photon energy pairs (not shown here) are generally as expected, showing a modest increase in the amplitude of the \EF\ peak relative to that of the 700~eV spectrum. The surface component, extracted from a photon energy pair of 60~eV and 700~eV, is essentially insulating and bears considerable resemblance to the AFI phase spectrum. This result raises the possibility of an insulating surface and a metallic bulk, as reported for other vanadium oxides\cite{Maiti98}.

The possible insulating behavior of the surface and the reduced \EF\ peak in less bulk sensitive low photon energy PES spectra are all due to the stronger correlation in the surface region. Since the atomic coordinate ``$z$" on the surface relative to that in the bulk is reduced, the bandwidth ``$zt$" on the surface is narrowed. The bandwidth reduction not only favors increased correlation by itself, but also tends to make the screening of $U$ be less effective and so results in a higher value of effective $U$ in the surface region. Within DMFT the reduction of the ratio of bandwidth to effective $U$ means that the QP peak at \EF\ becomes smaller\cite{HeldLong}, eventually resulting in a transition to an insulating system for a small enough ratio.

We note that the variation of the surface and bulk components is not exactly according to the phenomenological model given in Eq.~(\ref{SvsB_eq}). If we use pairs where the photon energies are more closely spaced, or with more closely spaced angles, or angle pairs including the 0 degree spectrum, we get results with greater variance, including sometimes a negative-going region for a surface component. However, the extracted bulk spectra always display both a -1.2~eV peak, which is interpreted as the lower Hubbard band (LHB), and the \EF\ peak, so that the former is certainly not absent in the correlated metal. Further, the difference between the bulk spectra and the 700~eV spectrum is not so great as to affect our discussion throughout the rest of this paper. Therefore, given the great uncertainties in extracting a bulk spectrum, we take the 700~eV spectrum to be an adequate representation of the bulk spectrum of \vo.

\subsection{\label{SS_SmallSpot}Importance of the small beam spot}

\begin{figure}[!t]
\includegraphics[width=3.0 in]{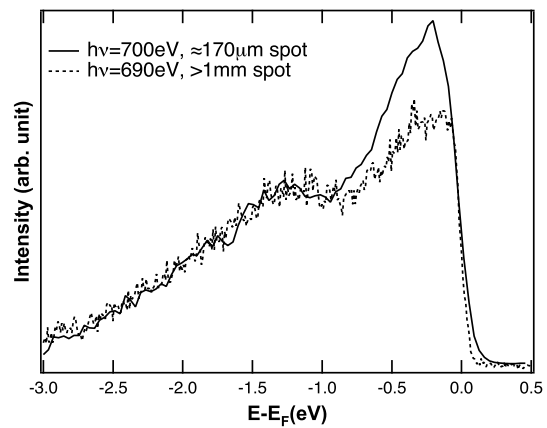}
\caption{\label{SmallSpot} Comparison of angle-integrated spectra taken with 1mm and $\approx$ 170$\mu$m beam spot diameters. The small difference in photon energy is not important for this comparison.}
\end{figure}

Our finding of the prominent \EF\ peak was only enabled by the use of a small beam spot size. Fig.~\ref{SmallSpot} compares two PES spectra taken with the same experimental setup except for different beam spot diameters, 1~mm and $\approx$ 170~$\mu$m. The spectra were taken at the same beamline at SPring-8 before and after an improvement in beam focusing, respectively. The difference in photon energy is too small to be responsible for the pronounced difference seen in the two spectra, that the \EF\ peak intensity is significantly lower for the larger spot size. Only for a spot of diameter $\approx$ 170~$\mu$m were we able to observe an intense \EF\ peak. This striking difference in spectra can be understood by considering the effect of steps and edges on the surface, as was demonstrated long ago for $f$-electron systems, both in theory\cite{Johansson} and experiment\cite{Domke86}. The reduction of the bandwidth from bulk to surface due to the reduced atomic coordination on the surface, explained in the previous subsection, immediately implies a further bandwidth reduction due to any departure from planar geometry that further reduces the coordination. Thus edges are more greatly affected than smooth surfaces and corners are more greatly affected than edges\cite{Johansson}. Indeed, the PES study of Ref.\ [\onlinecite{Domke86}] found new surface shifted lines in the spectrum from a surface that was deliberately roughened, in addition to the surface shifted peaks present for a smooth surface.

\begin{figure}[!bt]
\includegraphics[width=3.3 in]{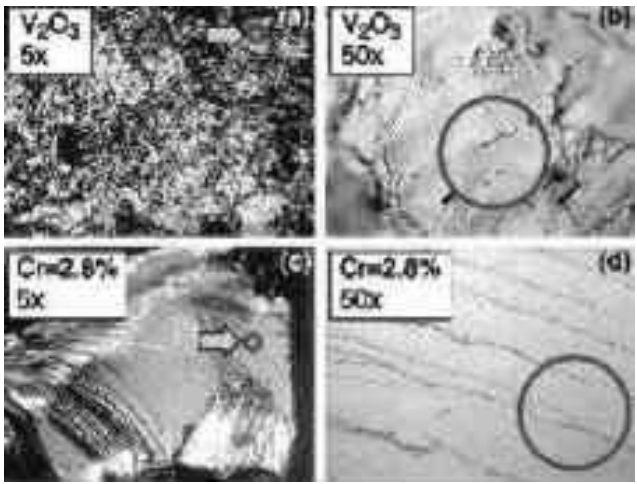}
\caption{\label{Microscope}(color online) Optical microscope figures of the surface of (a), (b) \vo\ and (c), (d) \vcrotwo, at 5x and 50x magnifications respectively. Circles correspond to a beam spot diameter of 100~$\mu$m, as at the ALS.}
\end{figure}

Fig.~\ref{Microscope} shows the optical microscope pictures of surfaces of (a), (b) \vo\ and (c), (d) \vcrotwo\ at 5x and 50x magnifications respectively. The circles in the figure represent the beam spot diameter of 100~$\mu$m, as at the ALS. One can clearly see that there exist many steps and edges on the sample surface, more so in \vo\ than in \vcrotwo. In general, we have observed that samples doped with Cr or Ti display more uniform cleaved surfaces than for the undoped ones, which might imply Cr or Ti doping stabilizes the surface of \vmo. But in general, for both \vo\ and \vcrotwo, compared to a scale of 1mm the number of steps and edges is much reduced on a scale of 100~$\mu$m. Thus we attribute the importance of the small beam spot to its ability to discriminate against these steps and edges. Indeed we have observed considerable variation of the \EF\ peak intensity with beam spot position when using the smaller spot size.

\subsection{Comparison to LDA+DMFT}

\begin{figure}[!tb]
\includegraphics[width=3.3 in]{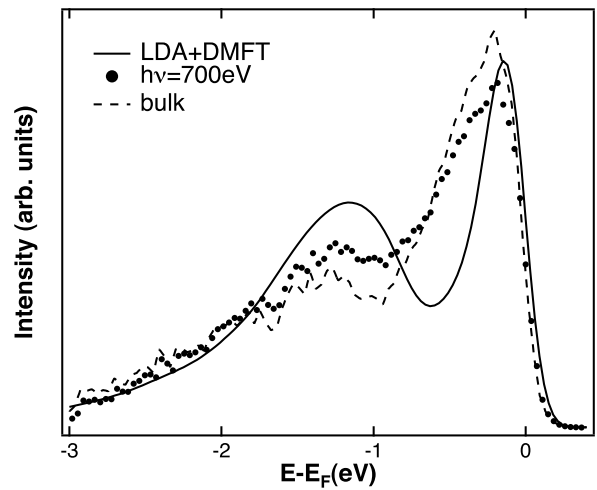}
\caption{\label{DMFT700eV} $h\nu$ = 700~eV spectrum of the PM phase \vo, along with a bulk spectrum from Fig.~\ref{Bulk_Surface}, compared with the LDA+DMFT (QMC) calculation made with $U$ = 5.0~eV and $T$ = 300~K. The theory curve is taken from Ref.\ [\onlinecite{Mo03}].}
\end{figure}

Fig.~\ref{DMFT700eV} reproduces from Ref.\ [\onlinecite{Mo03}] a comparison of the LDA+DMFT (QMC) calculation to the PES spectrum measured at $h\nu$ = 700~eV along with the bulk spectrum taken from Fig.~\ref{Bulk_Surface}. All three curves are normalized to have the same area, and the theory curve includes the Fermi function and Gaussian broadening of 90~meV to simulate the experimental resolution. As discussed in Ref.\ [\onlinecite{Mo03}], the theory and experiments are in general agreement in showing a pronounced QP peak near \EF\ and in having a LHB near -1.2~eV. However, the weight and the width of the experimental QP peak is significantly larger than in the theory. We reach the same conclusion even when an extracted bulk spectrum from Fig.~\ref{Bulk_Surface} is compared to the LDA+DMFT (QMC) calculation, with just a small change in the level of discrepancy of the spectral width and weight. 

The larger experimental width and weight imply weaker correlation that could be described by a reduced $U$ value in the DMFT. However, in the DMFT calculations made for \vmo\cite{Held01, Keller04, Laad03} so far, the MIT with Cr-doping would not occur for such a reduced $U$ value unless $U$ increases upon the transition into the PI phase. Another possible origin for the increased width of the experimental peak is the {\bf k}-dependence of the single particle self-energy, which is not included in DMFT calculations. This could be partially included by improved DMFT calculations, such as with the cluster DMFT recently applied to Ti$_2$O$_3$\cite{Poteryaev04}. The importance of a {\bf k}-dependent self-energy in understanding PES lineshapes has been reported for other strongly correlated transition metal oxides.\cite{Sekiyama97, Inoue95}. Also important is that the theory curve in Fig.~\ref{DMFT700eV} is calculated only for the V $3d$ $t_{2g}$ states, without including O $2p$ states, therefore missing a possibly important role of the hybridization between V $3d$ and O $2p$ in the electronic structure of \vmo. A theoretical scheme to calculate the whole valence band including O $2p$ is necessary to include this aspect. Very recently, a full orbital calculation scheme\cite{Anisimov05} using LDA+DMFT was suggested, incorporating O $2p$ states in the calculation of the electronic structure of \vo. However, a calculation made at low-$T$ enabling a direct comparison to our data is currently lacking. Another realization of the LDA+DMFT method\cite{Poteryaev06} that uses the Hamiltonian to construct the Green's function has been applied to \vo\ in a low temperature calculation and it finds improved agreement with the PM phase spectrum using a smaller value of $U$ than in the theory curve of Fig.~\ref{DMFT700eV}.

\section{\label{MIT}Metal-insulator transition}

\begin{figure}[tb]
\includegraphics[width=3.3 in]{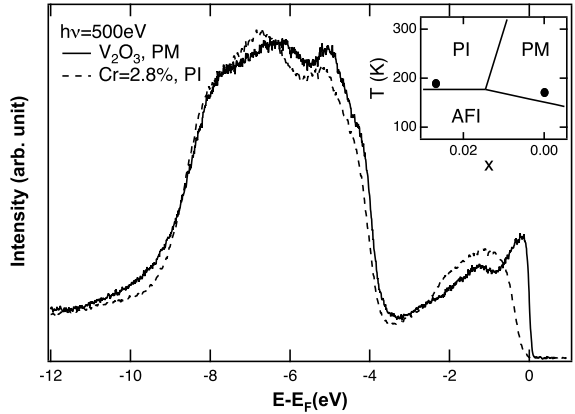}
\caption{\label{PM_PI_wide} PES spectra of the PM phase \vo\ and the PI phase \vcrotwo, taken with $h\nu$ = 500~eV, $T$ = 175~K (PM) and 190~K (PI). The inset shows the temperature and the Cr concentration of the data in a simplified phase diagram.}
\end{figure}

\subsection{PM-PI transitions}

Fig.~\ref{PM_PI_wide} shows the change in the PES spectrum for the PM to PI transition. The PM phase spectrum was taken on \vo\ at $T$ = 175~K and the PI phase spectrum was taken on \vcrotwo\ at $T$ = 190~K. Both spectra were measured with photon energy h$\nu$ = 500~eV. The O $2p$ and the V $3d$ parts of the spectrum are well separated in both phases, as already seen in Fig.~\ref{Overview} for the PM phase. In the energy range from -5~eV to -8~eV the two phases show some differences in the O $2p$ parts of their spectra. In the PM phase, one can also observe a weak structure at $\approx$ -9.5~eV, which has been explained within the Anderson impurity model as a V $3d$ satellite structure.\cite{Park_thesis} 

\begin{figure}[!bt]
\includegraphics[width=3.3 in]{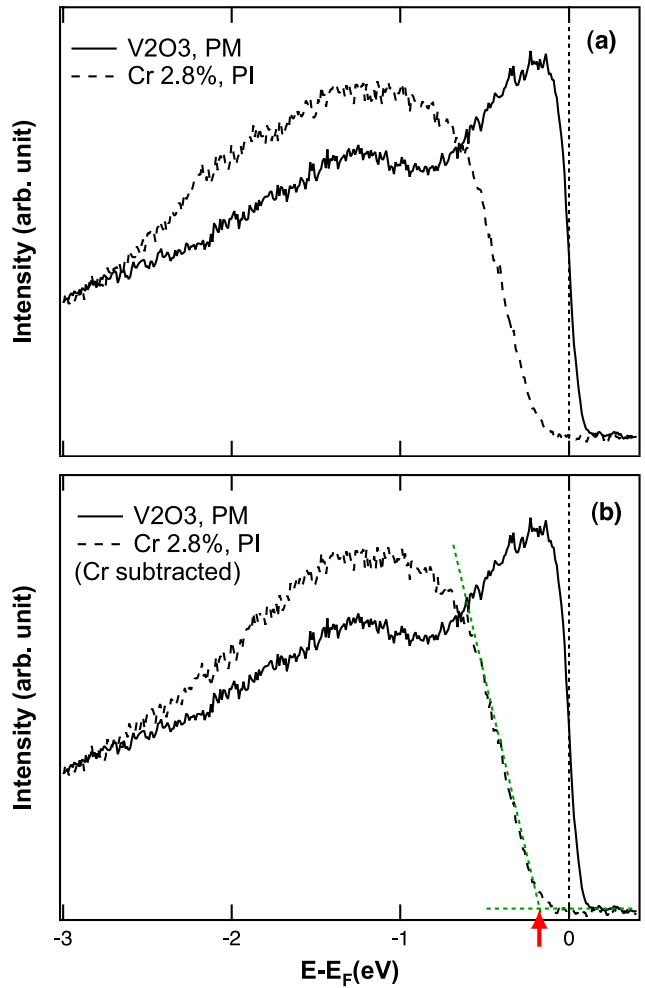}
\caption{\label{PM_PI_val}(color online) V $3d$ spectra of \vcro, taken with $h\nu$ = 500~eV, at $T$ = 175~K ($x$ = 0) and $T$ = 190~K ($x$ = 0.028). Small Cr emission around -2~eV seen in panel (a) is removed in panel (b) as explained in the text. The green lines and the red arrow in panel (b) are added to show the procedure of estimating the size of the gap as explained in the text. }
\end{figure}

Interestingly, the V $3d$ part of the PI phase spectrum is not a simple single peak, as can be seen better in Fig.~\ref{PM_PI_val}. The more complex structure of the insulating phase spectrum is even more clearly visible in the AFI phase spectrum, which will be presented in a later subsection. There are slope changes at $\approx$~-1.3 and $\approx$~-0.7~eV, giving the appearance that the coherent \EF\ peak is just pushed to higher binding energy when the MIT from the PM to PI phase occurs, with the LHB also being shifted to higher binding energy, but to a lesser degree. This apparent simple shift of the PM phase spectrum across the MIT is seen in a recent full-orbital DMFT calculation\cite{Anisimov05}, although the actual shape of that theory spectrum has differences from the experimental spectra, as discussed in the next subsection. The electron removal part of the PI phase gap is estimated as $\approx$~200~meV. The procedure of the estimation, as commonly used in semiconductor physics, is shown in panel (b). The two green lines are the base line of the spectrum and a straight line extrapolated from the leading edge of the PI spectrum. The difference in energy between the intersection of the two green lines (red arrow) and \EF\ is taken as the size of the gap. In so doing, we ascribe the tail of the spectrum to broadening due to the experimental resolution, thermal effects and other scattering processes. In fact, the tail of the spectrum is well reproduced from a model density of states that has a straight leading edge like the green line, by broadening with a Lorentzian of width $\lesssim$ 0.1 eV and Gaussian of width 0.1 eV. The typical error bar of the estimation is $\approx$ 25 meV. 

The spectrum of \vcro\ also shows a small peak centered at $\approx$ -2~eV. We identify this as emission from Cr $3d$, since the intensity of this peak is highly enhanced for a RESPES measurement made at the Cr \rt\ edge and the PES spectrum of Cr$_2$O$_3$ shows a Cr $3d$ peak centered around -2~eV\cite{Park_thesis}. Also, the intensity of this peak gets smaller in the spectrum of \vcroone\ at the same rate as the Cr concentration decreases. We have removed the Cr $3d$ emission, in Fig.~\ref{PM_PI_val}(b), by fitting the Cr emission part with a single Gaussian peak scaled by 0.28 and 0.12 for the PI phase spectra of \vcrotwo\ and \vcroone\ respectively, combined with the information obtained from the RESPES spectrum of \vcrotwo\ where the Cr emission is entirely suppressed, as described in the Appendix.

\subsection{Comparison of PI phase spectrum to DMFT}

\begin{figure}[!bt]
\includegraphics[width=3.3 in]{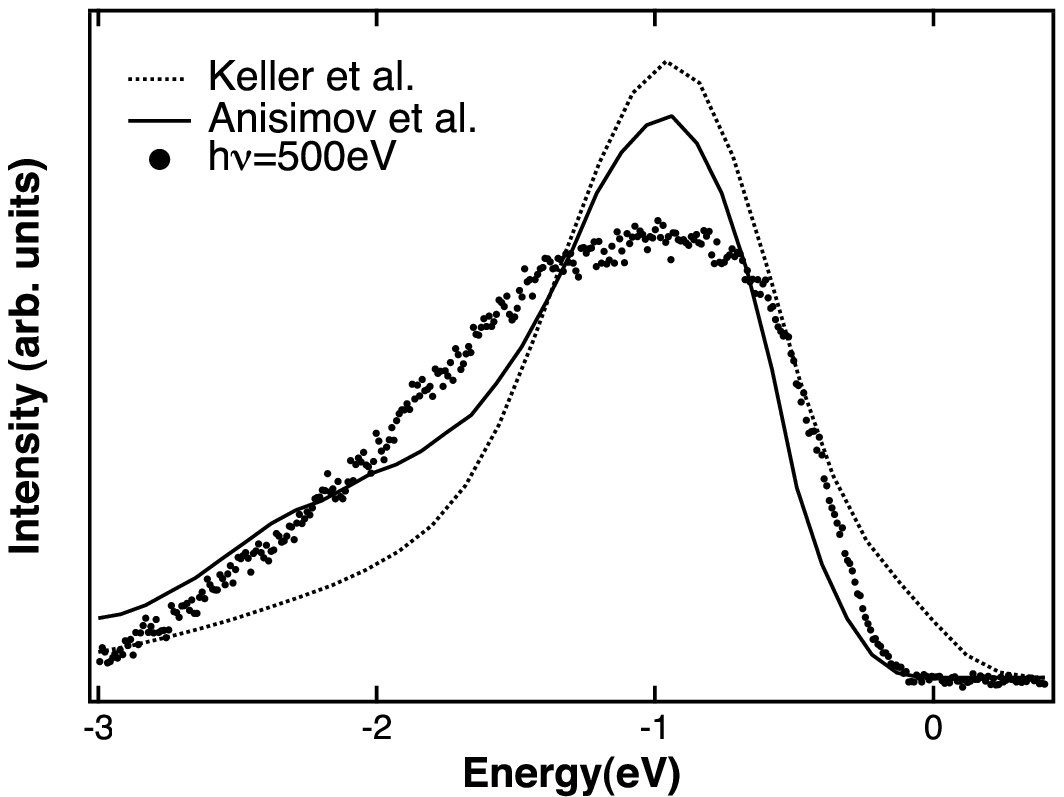}
\caption{\label{DMFT_PI} The PI phase spectrum of \vcrotwo, measured at $T$ = 190 K, compared to the LDA+DMFT calculations from Keller \etal\ (Ref.\ [\onlinecite{Keller04}]) and Anisimov \etal\ (Ref.\ [\onlinecite{Anisimov05}]).}
\end{figure}

We now turn our attention to the details of the PI phase spectrum by comparing it to spectra from LDA+DMFT calculations. Fig.~\ref{DMFT_PI} compares the PES spectrum of \vcrotwo, taken at $T$=190~K with $h\nu$ = 500~eV, with two different LDA+DMFT calculations. The Cr emission has been removed from the experimental data as described in the previous subsection. A Shirley-type background was subtracted from the experimental spectrum in an identical way as was done for the PM phase spectra, to facilitate the comparison. The curve labeled as Keller \etal\ is taken from Ref.\ [\onlinecite{Keller04}] and the one labeled as Anisimov \etal\ is from Ref.\ [\onlinecite{Anisimov05}]. The calculation for Keller \etal\ was made for $T$ = 700~K with $U$ = 5.0~eV using LDA+DMFT(QMC) for the $t_{2g}$ orbitals of (V$_{0.962}$Cr$_{0.038}$)$_2$O$_3$, and was broadened by a Gaussian of width 90~meV to simulate the experimental resolution. The small difference in the Cr concentration is not so important for our discussion here. The theory curve Anisimov \etal\ was calculated with $U$ = 5.5~eV and for $T$ = 1160~K using the full orbtal calculation scheme described in Ref.\ [\onlinecite{Anisimov05}]. Gaussian broadening of 0.2~eV was applied. 

The Keller \etal\ curve shows finite weight at the \EF, likely reflecting the fact that the calculation was made for $T$ = 700~K. Because this temperature is above the critical point of the phase diagram of \vmo, one might expect the building up of incoherent weight at \EF\cite{Bulla01, Limelette03} within the DMFT picture of the MIT. In fact, we have directly observed such a transfer of spectral weight into the insulator gap with increasing temperature\cite{Mo04a}. One may expect the suppression of this finite \EF\ weight when the calculation is made for a lower temperature, comparable to the experiment, analagous to the dramatic change with temperature of the LDA+DMFT calculation in the PM phase\cite{Mo03}. On the other hand, the curve of Anisimov \etal\ does not predict any left-over weight at \EF. It is worth noting that the main peaks of both theory curves are centered at $\approx$~-1~eV, the same as for the experimental spectrum. However, the theory curves fail to predict the more complex shape of the experimental spectrum, although the curve of Anisimov \etal\ does show a small secondary structure near -2~eV.

\subsection{PM-AFI transitions}

\begin{figure}[!tb]
\includegraphics[width=3.3 in]{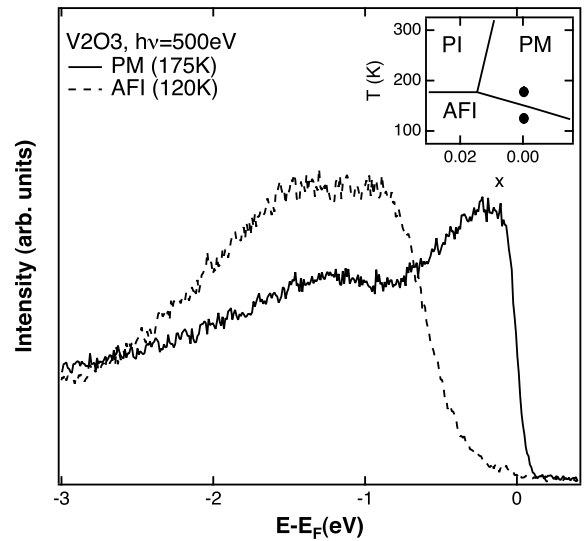}
\caption{\label{PM_AFI_pure} Spectral change across the PM to AFI transition in \vo. The inset shows the location of the data in the $T-x$ phase diagram of \vmo.}
\end{figure}

Fig.~\ref{PM_AFI_pure} shows the PM to AFI transition of the V $3d$ spectrum of \vo. The data were taken at $T$ = 175~K for the PM phase and at $T$ = 120~K for the AFI phase, as indicated also in the inset, with $h\nu$ = 500~eV. The spectra are normalized to have equal areas. The size of the electron removal gap is estimated in the same way as was done for the PI phase spectra of \vcrotwo, giving $\approx$~350~meV, which is 150~meV larger than that of the PI phase. More than for the PI phase, the AFI phase spectrum has somewhat the appearance of two peaks. This aspect of the data will be discussed more in the next subsection. 

\begin{figure}[!tb]
\includegraphics[width=3.0 in]{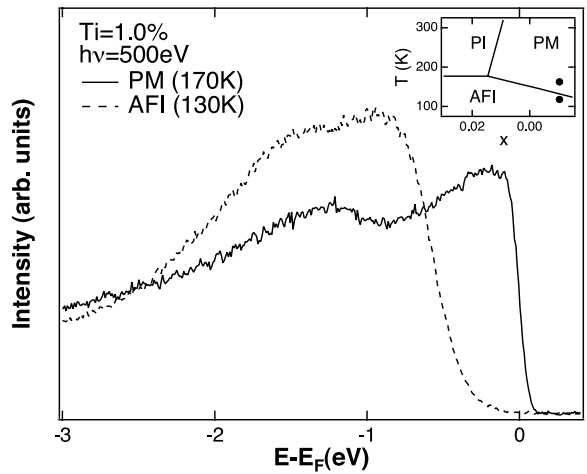}
\caption{\label{PM_AFI_Ti10} Spectral change across the PM to AFI transition in \vtioone. The inset shows the location of the data in the $T-x$ phase diagram of \vmo.}
\end{figure}

An almost identical change in the spectrum was observed for the PM-AFI transition of \vtioone\ as shown in Fig.~\ref{PM_AFI_Ti10}. These spectra were taken with $T$ = 170~K for the PM phase and $T$ = 130~K for the AFI phase with $h\nu$ = 500~eV. However, there are subtle but interesting differences compared to the PM and AFI phase spectra of \vo, as will be discussed in Sec.~\ref{Doping}.

\subsection{MIT in \vcroone\ }
\begin{figure}[!bt]
\includegraphics[width=3.3 in]{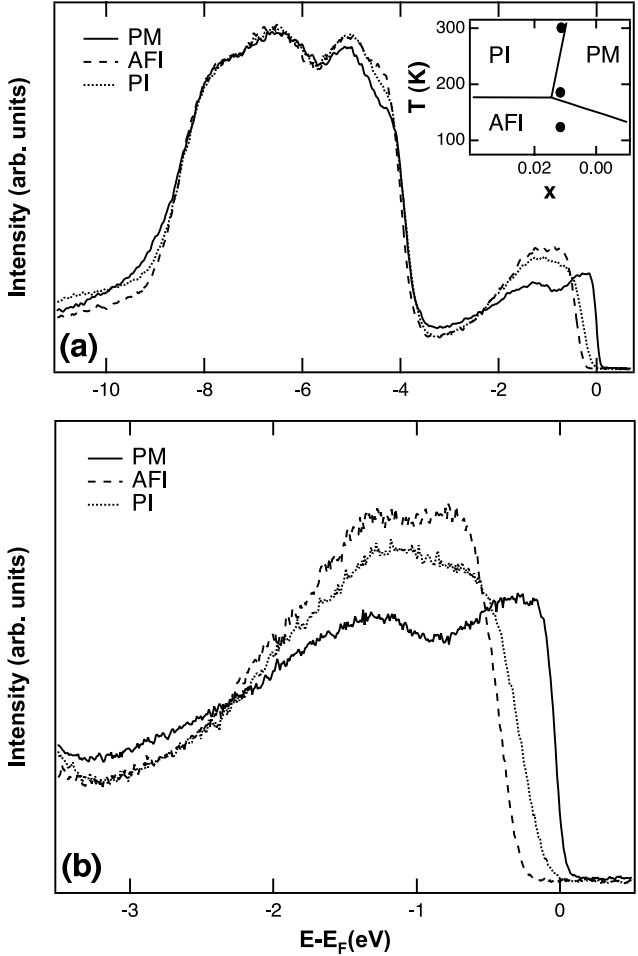}
\caption{\label{Cr12} MIT in \vcroone. For this Cr concentration, one can reach all three phases of \vmo\ just by changing the temperature. The spectra are presented for (a) a wide energy range including the O $2p$ part and (b) near \EF\ for V $3d$ part. The inset in (a) shows the locations of the data in the $T-x$ phase diagram of \vmo.}
\end{figure}

\vcroone\ has a composition that is special because one can reach all three phases of \vmo\ with a single sample just by changing the temperature. Fig.~\ref{Cr12} summarizes the data taken for \vcroone, at $T$ = 120~K (AFI), $T$ = 190~K (PM), and $T$ = 310~K (PI), with $h\nu$ = 500~eV. In the overview spectra shown in panel (a), it is interesting to note the differences for the different phases in the O $2p$ parts of the spectra. Different from what is observed in the PM-PI transition from \vo\ to \vcrotwo, the change occurs primarily for the leading structure around -5~eV, while the peak near -6.5~eV shows hardly any change. Therefore, we may ascribe the change in the -6.5~eV structure seen in Fig.~\ref{PM_PI_wide} mostly to the Cr-doping, while attributing the change in -5~eV structure mainly to the phase transitions. It is also noteworthy that the satellite peak at $\approx$ -9.5~eV is still visible in the PM phase of \vcroone\, the same as was found in the PM phase of \vo. The flat part near -9.5~eV in the PI phase has a higher intensity than in the AFI phase, while the spectral shape is nearly identical. Since we did not find any satellite peak in other PI phase spectra and the measurement temperature is relatively higher than that of the AFI phase spectrum, this intensity should be attributed to a higher inelastic background for this sample, rather than to an actual satellite peak. 

Panel (b) gives a view of the V $3d$ spectra of all three phases of \vmo. The data are taken from the same surface as that in Fig.~\ref{Cr12}(a), with a higher resolution. From the PI phase to the AFI phase, the leading edge is shifted by $\approx$~150~meV, consequently increasing the size of the gap by essentially the same amount as deduced by comparing the AFI and PI phase spectra from samples with different $x$-values in the preceding subsection. More interestingly, the spectral weight pushed to higher binding energy by widening the gap accumulates mostly in the region near \EF\ in the AFI phase, enhancing the appearance of two peaks for the AFI phase relative to that for the PI phase. The increase of the size of the gap through the PI to AFI transition has been predicted from the $t$-$J$ model in the large $U$ limit.\cite{Metzner92} A recent DMFT study\cite{Sangiovanni05} made for the one-band Hubbard model also predicts such an increase of the gap size. In the same study, the peak at the upper edge of the AFI phase spectrum is identified as the part of the spectrum arising from a renormalized Slater band quasi-particle. A more detailed presentation of this calculation and the comparison to experiment can be found in Ref.~[\onlinecite{Sangiovanni05}].

\section{\label{Doping}Spectral change with doping}

\begin{figure}[!bt]
\includegraphics[width=3.3 in]{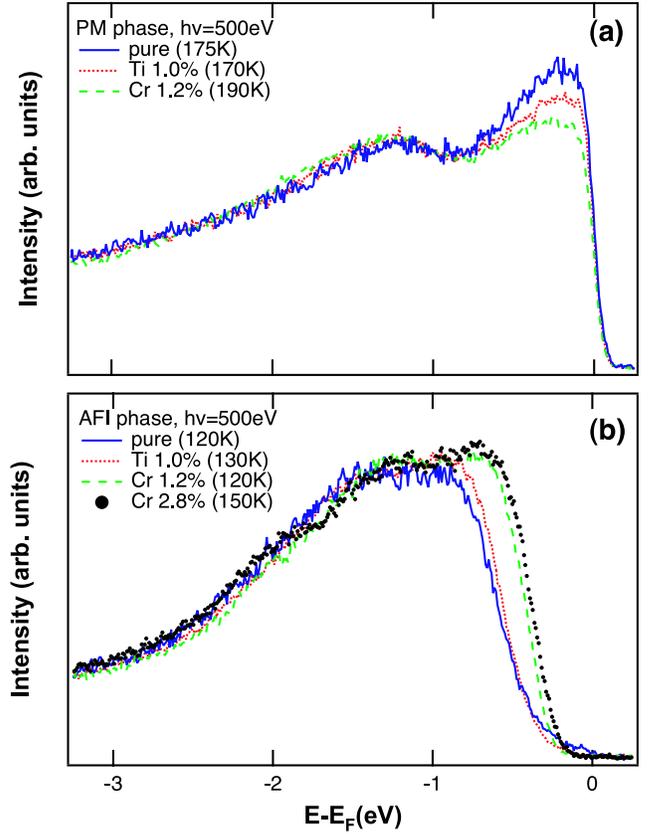}
\caption{\label{xDiff}(color online) Difference in the lineshapes as the dopant concentration ``$x$" changes, in (a) the PM and (b) the AFI phase.}
\end{figure}

In the previous sections, we have concentrated on the spectral changes across the phase boundaries. While the general features of these changes are very similar for the various samples, there are subtle but interesting differences in the lineshapes of the spectra of different samples within a given phase.

Panel (a) of the Fig.~\ref{xDiff} shows the PM phase spectra of \vo, \vcroone, and \vtioone\ taken at very similar temperatures, $T$ = 175~K, 190~K, and 170~K respectively. All the data were taken with 500~eV photon energy, and they are normalized to the height of the -1.2~eV peak, rather than to equal areas, to ease the comparison of the \EF\ peaks. The spectra show a monotonic decrease in the peak height as the concentrations of the dopants increase, without respect to their kind, i.e., Cr or Ti. The rate of the decrease in the \EF\ peak height is not a simple linear function of the dopant concentration, so the change from \vo\ to \vtioone\ is nearly the same as the change from \vtioone\ to \vcroone. It is not easy to determine if the decrease in peak height reflects an increase of width, or a decrease of spectral weight or both. Qualitatively the decrease in height correlates with the change of the resistivity \cite{Kuwamoto80, Carter91}, in that for the dopants and temperatures of the data reported here, the resistivity $\rho$ increases in the same sequence. However, the rate of increase in $\rho$ is not quite in accordance with the decrease in the peak height. The resistivity increases by less than 2 times when 1.0\% Ti is doped to \vo, while it increases more than an order of magnitude with 1.2\% Cr-doping.

The spectral change in the AFI phase is quite different from that of the PM phase, as can be seen in panel (b) of Fig.~\ref{xDiff}. As the doping level increases from \vo\ to \vtioone\ to \vcroone\ to \vcrotwo, the high binding energy part of the spectrum stays nearly the same, but the peak closest to the chemical potential becomes larger and its onset shifts to lower binding energy, i.e., the AFI phase gap decreases. For \vo\ and \vtioone, the electron removal gap is $\approx$ 350 meV, and it becomes $\approx$ 270 meV and 230 meV for \vcroone\ and \vcrotwo, respectively. In the framework of LDA+DMFT for the one-band Hubbard model\cite{Sangiovanni05} one would say that the renormalized Slater peak becomes more pronounced upon moving deeper into the AFI phase for Cr 2.8\%, which may be plausible. However, one might also expect that the gap size would increase, contrary to what is seen.

\section{\label{Summary}Summary}

We have measured the photoemission spectra of \vo, \vcroone, \vcrotwo, and \vtioone\ in the PM, PI, and AFI phases. All our data were taken with photon energies that are away from that for V $3d$ RESPES, to avoid Auger emissions that distort the lineshapes as described in the Appendix. In the PM phase of \vo, we have observed a prominent QP peak, which can only be observed with the combination of high photon energy to increase bulk sensitivity and a small beam spot to reduce the effects of surface steps and edges. The observed spectra in the PM phase of \vo\ are in general agreement with an LDA+DMFT(QMC) calculation, but show a significantly larger peak width and weight. Across the PM-PI and PM-AFI phase boundaries, the V $3d$ parts of the spectra in the insulating phases show structures that are more complex than a single peak, more clearly so in the AFI phase. 

The size of the electron removal gap increases as the PI to AFI transition occurs, and the peak close to \EF\ intensifies. For example, from the PI phase to the AFI phase of \vcroone, the gap size increases from $\approx$ 120 meV to $\approx$ 270 meV. The spectra taken in the same phases with different compositions show interesting, but not explained, monotonic changes as the dopant level increases, regardless of the dopant species. In the AFI phase, the size of the gap decreases as the dopant level increases. It is $\approx$ 350 meV for \vo\ and \vtio, and it decreases to $\approx$ 270 meV and $\approx$ 230 meV for \vcroone\ and \vcrotwo, respectively. In contrast, for the PI phase, the gap size {\em increases} from $\approx$ 120 meV to $\approx$ 200 meV for the two Cr dopings, respectively. That is, the jump in gap across the PI to AFI transition {\em decreases} for the two Cr dopings, respectively.

\begin{acknowledgments}
This work was supported by the U.S. NSF at the University of Michigan (Grant No.~DMR-03-02825), by the U.S. DoE at the ALS (Contract No.~DE-AC03-76SF00098), by the Center for Strongly Correlated
Materials Research, Korea, by Grant-in-Aid for COE Research(10CE2004) and Grant-in-Aid for Creative Scientific Research(15GS0213) of MEXT, Japan, by 21st Century COE Program(G18) by Japan Society for the Promotion of Science, by KOSEF through eSSC at POSTECH. 
We are very grateful to K. Held, D. Vollhardt, G. Keller, V. Anisimov, and G. Sangiovanni for valuable discussions.
\end{acknowledgments}

\

\appendix*
\section{Resonant photoemission and Auger emission}

\begin{figure*}[tb]
\includegraphics[width=7.0 in]{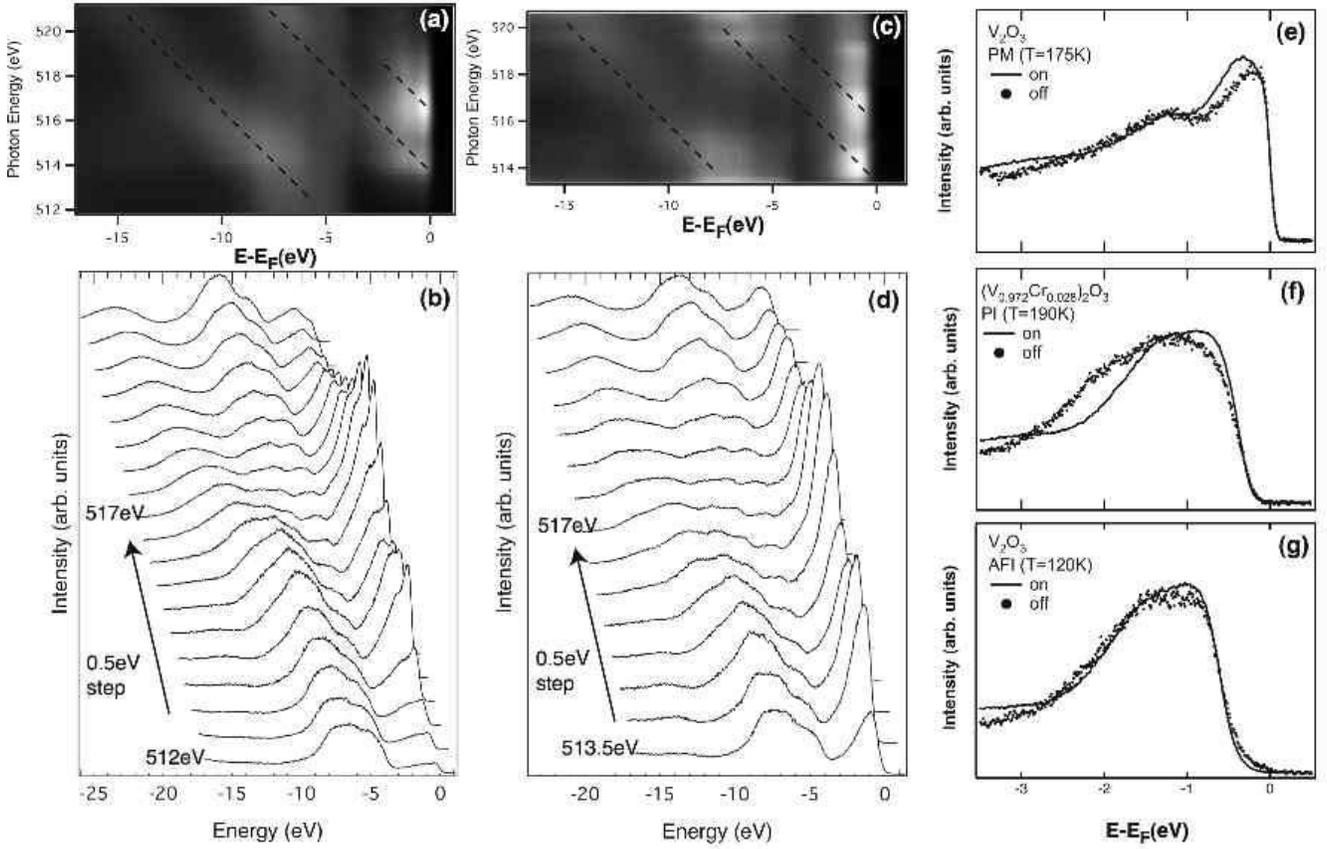}
\caption{\label{ResPES}RESPES spectra of \vo\ in (a),(b) the PM phase at $T$ = 185~K and (c),(d) the AFI phase at $T$ = 160~K, taken at the ALS. (a) and (c) are intensity maps generated from the PES spectra shown in (b) and (d). The black dotted lines in (a) and (c) are added to guide the eyes to the Auger peaks that are seen at constant kinetic energies. The distortion of the V $3d$ lineshapes close to \EF\ in on-resonance PES spectra is shown in (e) - (g) for the PM, PI, and AFI phases, respectively. The data in (e) - (g) were taken at SPring-8 with $h\nu$ = 500 eV for off-resonance spectra and $h\nu\ \approx$ 517 eV for on-resonance spectra. Panel (e) is from Ref.\ [\onlinecite{Mo04b}].}
\end{figure*}

The RESPES spectra of \vmo\ at the V \rt\ edge and the accompanying Auger emission that we have found are summarized in Fig.~\ref{ResPES}. Panels (a) and (b) show the PM phase RESPES spectra of \vo\ measured at the ALS, with 0.5 ~eV steps from 512 ~eV to 521 ~eV photon energy, (a) in an intensity map and (b) as stacked PES spectra. The temperature for the measurement was 185~K. The energy axis of panel (b) is labeled for the 512~eV spectrum, and other spectra are shifted by multiples of 0.5~eV to higher binding energy to enhance the visibility. All the spectra are normalized to the incident photon current, so that their relative intensities have physical meaning. The resonant enhancement of the V $3d$ valence band spectrum is clearly seen with the maximum for $h\nu \approx\ $517 ~eV. The dashed lines in panel (a) are guidelines to show the Auger peaks, occurring at constant kinetic energies, and therefore moving by 0.5 ~eV to higher binding energy as the photon energy increases by 0.5 ~eV. The Auger peaks that center around -11 ~eV and -3 ~eV in the 517 ~eV spectrum show this constant kinetic energy behavior very clearly, while that of the near \EF\ emission is less visible in both panel (a) and (b). This weak emission can only be seen in the high-resolution on- and off- spectra discussed further below.

We have also observed similar Auger emission in the RESPES spectra of the AFI phase of \vo, as shown in panels (c) and (d). The spectra are taken at the ALS with $T$ = 160~K and photon energies ranging from 513.5 ~eV to 520.5 ~eV. They are presented in the same way as in panels (a) and (b). One can identify the constant kinetic energy peaks that fall near -11 and -3~eV in the 517~eV spectrum just as in the PM phase spectra.

Higher resolution data, taken for the PM phase of \vo, the PI phase of \vcrotwo, and the AFI phase of \vo\ are shown in panels (e), (f), and (g) respectively. Each panel compares on- and off-resonance spectra. All spectra were taken at SPring-8 with a resolution of $\approx$~90~meV, and are normalized to the intensity around -1.2 ~eV to enhance the visibility of the comparison of the lineshapes close to \EF. Small Cr-emission is clearly visible in the off-resonance spectrum in panel (f), but it is completely suppressed in the on-resonance spectrum. The on-resonance spectra in all three panels show Auger emission centered around -3~eV, which is already seen in panel (a) - (d). In addition, the on-resonance spectrum of the PM phase shows an extra feature close to \EF\ peaks around -0.2~eV, and therefore yielding a different lineshape from that of the off-resonance spectrum. In the insulating phases, the Auger emission causes an increase of the peak intensity close to the \EF, compared to that of the off-resonance spectra. 

For the detailed comparison to theory, especially in the PM phase, the distortion of the lineshape resulting from the Auger emission was significant enough for us to deliberately avoid the resonance region in choosing the photon energy for our measurements. All the spectra presented in the main body of this paper were measured using photon energies away from the resonance region.



\begin{thebibliography}{99}
\bibitem{footnote_beamspot} We previously\cite{Mo03} underestimated the beam spot diameter in the work at SPring-8 to be $\approx$ 100 $\mu$m.
\end{thebibliography}
\end{document}